\documentclass[superscriptaddress,prc,preprint,showpacs]{revtex4}

\date{\today}
\usepackage{graphicx}
\usepackage{amssymb}
\usepackage{amsmath}
\usepackage{dcolumn}

\newcommand{\calcium}[1][40]{{}^{#1}\mathrm{Ca}}
\newcommand{\iron}[1][]{{}^{#1}\mathrm{Fe}}

\newcommand{\MSU}{National Superconducting Cyclotron Laboratory and
  Department of Physics and Astronomy, Michigan State University, East
  Lansing, Michigan 48824}
\newcommand{\WIS}{
  Department of Physics, University of Wisconsin,
  Madison, WI 53706, USA}
\newcommand{\Tohoku}{Department of Physics, Tohoku University, Sendai
  980-8578, Japan}

\begin{document}

\title{Symmetry energy for fragmentation in dynamical nuclear
collisions}

\author{Akira Ono}
\affiliation{\Tohoku}
\author{P. Danielewicz}
\affiliation{\MSU}
\author{W. A. Friedman}
\affiliation{\WIS}
\author{W. G. Lynch}
\affiliation{\MSU}
\author{M. B. Tsang}
\affiliation{\MSU}

\begin{abstract}
  We extract values for the free symmetry energy as a function of the
  fragment size (the proton number $Z$) from antisymmetrized molecular
  dynamics (AMD) calculations of calcium collisions.  Simple
  statistical physics describe well the distribution of hot nuclei at
  breakup, provided the surface symmetry term in the free energy is
  much smaller at high excitation than in ground state nuclei. This
  result may reflect the condition of low density and  finite
  temperature when these systems disassemble.
\end{abstract}

 \maketitle

%\section{Introduction}

Free energies play an important role in mixed phase environments.
For nucleonic systems, such environments may be found in
subsaturation density systems formed in nucleus-nucleus
collisions, supernova collapses \cite{bethe} and in the inner
crusts of neutron stars \cite{pethick,lattimer-prakash}. In these
scenarios, the free energy of fragments largely defines the
balance between denser fragments and the more dilute nucleonic
gas. Key uncertainties to the prediction of free energies are
their dependencies on temperature, density and isospin asymmetry.
The latter uncertainty is particularly relevant to the inner crust
of neutron stars, where the nuclear isospin asymmetry term
stabilizes nuclear droplets enveloped by a lower density neutron
gas \cite{pethick,lattimer-prakash}. The development of experiment
constraints on the asymmetry term constitutes an important
scientific objective.

Here we focus on multifragmentation in nucleus-nucleus collisions.
Many aspects of such data have been successfully described by
equilibrium statistical models \cite{gross,bondorf,tan} that assume
fragments to be produced by a low density phase transition that occurs
during the expansion stage of a collision. The success of such
descriptions suggests applications of statistical theory to such
collisions may provide constraints on the free energies and their
asymmetry dependence. Such comparisons provide an experimental test of
the validity of equilibrium descriptions. One can also test the
assumption of local thermal equilibrium by comparing the predictions
of modern transport theories to equilibrium models. Calculations of
isotopic observables are particularly sensitive to the degree of
chemical equilibrium achieved.

We investigate these issues with the antisymmetrized molecular
dynamics (AMD) code of Ref.\ \cite{ONOi}.  In the present work, we
analyse the yields of fragments produced in AMD simulations for
central collisions of nuclei with $40\le A\le 60$  at an incident
energy $E/A=35$ MeV. Even though AMD does not require the
assumption of equilibrium, we have shown in a previous paper
\cite{ONOk} that some observables obtained from predicted isotope
yields, such as isoscaling,  are consistent with the expectations
from a statistical interpretation of fragment formation.

If the yield of fragments, produced with $N$ neutrons and $Z$
protons in these collisions, is governed by a statistical process
at constant pressure, the fragment yield $Y(N,Z)$ can be related
to the nuclear free energy $G_\text{nuc}(N,Z)$ by the relation
\begin{equation}
Y(N,Z)\propto\exp\biggl[-\frac{G_\text{nuc}(N,Z)}{T}
                  +\frac{\mu_n}{T}N+\frac{\mu_p}{T}Z\biggr],
\label{eq:YNZequi}
\end{equation}
where $\mu_n$ and $\mu_p$ are the neutron and proton chemical
potentials.  At temperatures relevant for  multi-fragmentation,
the temperature dependence of the asymmetry term in the free
energy for bulk nuclear matter is expected to be negligible.
However, the temperature dependence of the surface symmetry term
of the free energy is not well known. The present study suggests
that the surface contribution to the symmetry free energy may be
weakened under the conditions associated with fragment formation.

\begin{figure*}
\includegraphics[width=\textwidth]{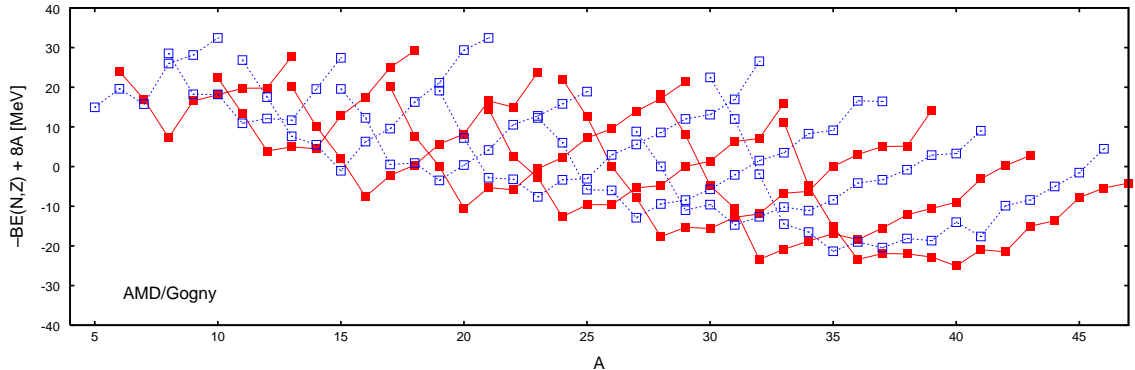}
\caption{\label{fig:mass-gogny}Binding energies (not per nucleon)
  of nuclei calculated by AMD with the Gogny force.  The quantities
  $-\mathrm{BE}(N,Z)+8(N+Z)\ \mathrm{MeV}$ are shown by filled and
  open squares for even-$Z$ and odd-$Z$ nuclei, respectively.  Lines
  connect isotopes.}
\end{figure*}

Calculations within the AMD formalism have demonstrated its capability
of reproducing many quantal features of ground state nuclei
\cite{AMD-Structure}. The present AMD code was streamlined for
collisions; it nevetheless reproduces ground state binding energies of
nuclei to within about 0.5 MeV/nucleon. Figure \ref{fig:mass-gogny}
shows the binding energies of nuclei $\mathrm{BE}(N,Z)$ with
$A\lesssim 40$ which are obtained by minimizing the energy within AMD
by adopting the Gogny force \cite{GOGNY}. In order to remove the
strong $A$-dependence, we plot $-\mathrm{BE}(N,Z)+8A$ MeV. We note
that the free energy equals the energy itself at zero temperature and
pressure, i.e.  $G_\text{nuc}(N,Z,T=0,P=0)=-\mathrm{BE}(N,Z)$.

Much of our present knowledge of the symmetry energy has been
extracted by fitting the nuclear binding energies with the
liquid-drop mass formula which contains a symmetry energy term
such as
\begin{equation}
E_\text{sym}(N,Z)=c(A)\frac{(N-Z)^2}{A}. \label{eq:symengterm}
\end{equation}
In the simplest mass formula \cite{WEIZSACKER}, the coefficient is
independent of the nuclear size, $c(A)=c_\text{sym}$, which
assumes the volume nature of the symmetry energy.  On the other
hand, advanced mass formulas
\cite{MYERS-SWIATECKI,MOLLER-NIX,DANIELEWICZ-symeng} have
introduced the $A$-dependence of $c(A)$ as the surface effect. The
extraction of $c(A)$ from the nuclear binding energies is not
quite straightforward even for ground state nuclei.  For example,
if one extracts the symmetry energy by using the energy difference
of neighboring nuclei \cite{JANECKE}, the symmetry energy is
largely a fluctuating function in the nuclear chart due to shell
and paring effects.  Nevertheless, the global fitting with the
assumption of $c(A)=c_\text{v}+c_\text{s}A^{-1/3}$, together with
the standard volume, surface, Coulomb and paring terms, usually
results in a reasonable value of the coefficients.  For example,
we obtain $c_\text{v}=27.3$ MeV and $c_\text{s}=-23.7$ MeV by
fitting all the available information of the binding energies of
nuclei \cite{AME}, while we obtain $c_\text{v}=30.1$ MeV and
$c_\text{s}=-35.1$ MeV by fitting the nuclei with $7\le A\le 40$.

If we fit the AMD binding energies in Fig.\ \ref{fig:mass-gogny}
for $7\le A\le 40$,  we obtain $c_\text{v}=30.9$ MeV and
$c_\text{s}=-35.2$ MeV.  The extracted value of $c_\text{v}$ is
comparable with the symmetry energy in infinite nuclear matter at
saturation density $\rho_0$.  To a good approximation, the nuclear
matter EOS at zero temperature can be written as
\begin{equation}
E(\rho,\delta)/A= E(\rho,\delta=0)/A+C_\text{sym}(\rho)\delta^2,
\end{equation}
where $\rho=\rho_n+\rho_p$ and $\delta=(\rho_n-\rho_p)/\rho$, and
the symmetry energy coefficient has the value
$C_\text{sym}(\rho_0)=30.7$ MeV for the Gogny force.  The present
study of heavy ion fragmentation reactions will provide
information about the symmetry free energy at subsaturation
densities and finite temperatures.

In order to extract the symmetry free energy from Eq.\
(\ref{eq:YNZequi}), it is necessary to obtain the fragment yields
$Y(N,Z)$ for a wide range of isotopes.  This is practically impossible
if we utilize the products produced in only one reaction
system. However, we overcome this limitation by combining the results
from various reaction systems here labeled by an index $i$.  We
calculate $\calcium[40]+\calcium[40]$ ($i=1$),
$\calcium[48]+\calcium[48]$ ($i=2$), $\calcium[60]+\calcium[60]$
($i=3$) and $\iron[46]+\iron[46]$ ($i=4$) collisions at zero impact
parameter and an incident energy $E/A=35$ MeV. Previous calculations
with an equivalent version of AMD and the Gogny force
\cite{ONOh,WADAa} reproduce experimental data for various fragment
observables in $\calcium[40]+\calcium[40]$ collisions at $E/A=35$ MeV.

AMD represents the wavefunction of the colliding system by fully
antisymmetrized products of Gaussian nucleon wave packets and
propogates these wave packets microscopically during the collision
\cite{ONOab,ONOh,ONOi}. The centroids of the nucleonic wave
packets move deterministically through the mean field potential
formed by the interactions with other nucleons. In addition, the
followed state of the simulation branches stochastically and
successively into a huge number of reaction channels. The
branching is caused by the two-nucleon collisions and by the
splittings of the wave packets. The interactions are parameterized
in the AMD model in terms of an effective force acting between
nucleons and in terms of the two-nucleon collision cross sections.
We adopt the Gogny effective force \cite{GOGNY} in this paper.

We simulate collisions by boosting two nuclei whose centers were
separated by 9 fm and calculating the dynamical evolution of each
collision until $t=300$ fm/$c$. The numbers of simulated events are
1040, 949, 978 and 1400, respectively, for the four systems.  In
central collisions, as shown in previous papers \cite{ONOh,WADAa}, the
projectile and target basically penetrate each other and many
fragments are formed not only from the projectilelike and targetlike
parts but also from a neck region between the two parts.  The nuclear
matter seems to be strongly expanding one-dimensionally in the beam
direction.

As we have shown in Ref.\ \cite{ONOk}, the AMD results satisfy
the isoscaling relation \cite{XU-isoscale}
\begin{equation}
Y_{i}(N,Z)/Y_{i'}(N,Z)\propto e^{\alpha N+\beta Z}\label{eq:isoscaling}
\end{equation}
with parameters $\alpha$ and $\beta$, for the fragment yields from two
different reaction systems ($i$ and $i'$) with different
neutron-to-protion ratios.  When we have four reaction systems
($i=1,2,3,4$), the fragment yields $Y_i(N,Z)$ are related to the
yields $Y_1(N,Z)$ of the reference system by
\begin{equation}
Y_1(N,Z)=Y_i(N,Z)e^{-\alpha_i N-\gamma_i(Z)}.
\label{eq:yone}
\end{equation}
The yields $Y_1(N,Z)$ for one reaction system can have good statistics
only in a narrow region of $(N,Z)$.  The isoscaling relation is
utilized with respect to $N$ to extend the range of $(N,Z)$.  We take
the $\calcium[40]+\calcium[40]$ system as the reference system and
therefore $\alpha_1=\gamma_1(Z)=0$.  The quantity $-\ln Y_1(N,Z)$,
i.e.\ the fragment yields for $\calcium[40]+\calcium[40]$, can be
represented in four different ways by using the scaled yields for
different reaction systems which have good statistics in different
regions.  They are combined by
\begin{align}
K(N,Z)
&=\sum_{i=1}^4
 w_i(N,Z)[-\ln Y_i(N,Z)+\alpha_iN+\gamma_i(Z)]\label{eq:KNZaverage}
\end{align}
where the averaging weights $w_i(N,Z)$ are determined by minimizing
the statistical errors in $K(N,Z)$ for individual $(N,Z)$.  The
isoscaling parameters $\alpha_i$, which is common to all $Z$, have
been obtained by the isoscaling fit \cite{ONOk}.  The parameters
$\gamma_i(Z)$ ($i=2,3,4$) for each $Z$ are determined by optimizing
the agreement of the quantities $[-\ln Y_i(N,Z)+\alpha_i+\gamma_i(Z)]$
from different reactions $i=1,2,3$ and 4.

\begin{figure*}
\includegraphics[width=\textwidth]{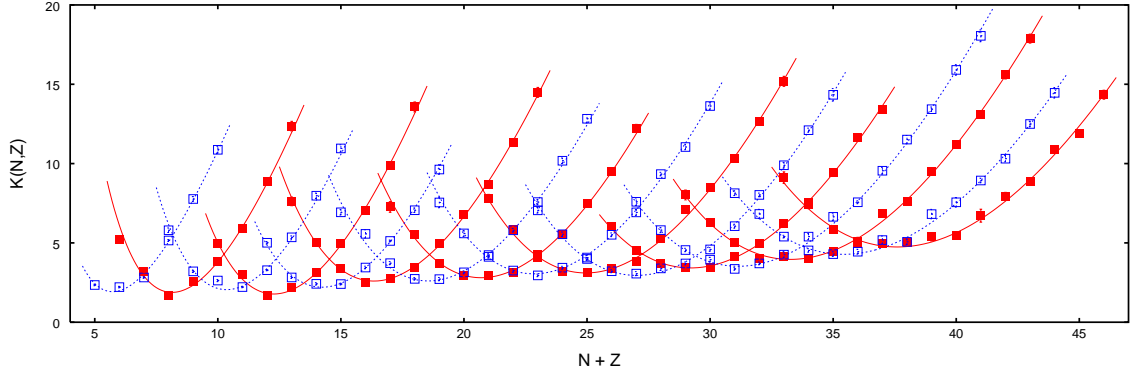}
\caption{\label{fig:logyfit-fit} The values of $K(N,Z)$ for $3\le Z
  \le 18$ are shown by symbols for the abscissa of $N+Z$.  The values
  are obtained by combining the results of
  $\calcium[40]+\calcium[40]$, $\calcium[48]+\calcium[48]$,
  $\calcium[60]+\calcium[60]$ and $\iron[46]+\iron[46]$ simulations.
  The error bars show the statistical uncertainty due to the finite
  number of events.  The curve for each $Z$ was obtained by fitting
  $K(N,Z)$ using Eq.\ (\ref{eq:KNZfit}).}
\end{figure*}

The values of $K(N,Z)$ obtained from the AMD simulations are shown by
the open and filled squares in Fig.\ \ref{fig:logyfit-fit} for odd-
and even-$Z$ nuclei, respectively.   Even though the event numbers are
not very large, $K(N,Z)$ has been obtained for a wide region of
$(N,Z)$.  In most cases $K(N,Z)$ covers more than 10 isotopes with
good statistical precisions for each $Z$.

%% \begin{figure}
%% \includegraphics[width=0.45\textwidth]{logyfit-deltazoasq.ps}
%% \caption{\label{fig:logyfit-deltazoasq} Difference of $(Z/A)^2$ of
%% fragments between $\calcium[40]+\calcium[40]$ and
%% $\calcium[60]+\calcium[60]$ collisions as a function the fragment
%% charge $Z$.  This is the result of the AMD simulations at $t=300$
%% fm/$c$ with Gogny force (open squares) and Gogny-AS force (solid
%% circles).}
%% \end{figure}

So far, we have not assumed any specific form of  $K(N,Z)$.
Nevertheless, the results show a very smooth behavior of $K(N,Z)$
as a function of $N$ and $Z$.  The shell and paring effects are
weak in $K(N,Z)$ compared to the ground state binding energies
shown in Fig.\ \ref{fig:mass-gogny}.  Each curve in Fig.\
\ref{fig:logyfit-fit} shows the fitting of $K(N,Z)$ for each $Z$
by a function
\begin{equation}
K(N,Z)=\xi(Z)N+\eta(Z)+\zeta(Z)\frac{(N-Z)^2}{N+Z},
\label{eq:KNZfit}
\end{equation}
where $\xi(Z)$, $\eta(Z)$ and $\zeta(Z)$ are the fitting parameters.
The result of the simulations is fitted well by this functional form.
We choose the quadratic term similar to Eq.\ (\ref{eq:symengterm}) for
convenience, so that the parameter $\zeta(Z)$ is directly related to
the symmetry energy as shown below.

%% In our previous work \cite{ONOk}, by using Eq.\
%% (\ref{eq:KNZfit}), the fragment isospin asymmetry is related to the
%% isoscaling parameter by a linear relation
%% \begin{equation}
%% \frac{\alpha_i-\alpha_1}{[Z/\bar{A}_1(Z)]^2-[Z/\bar{A}_i(Z)]^2}=4\zeta(Z),
%% \label{eq:linearrel}
%% \end{equation}
%% where $\bar{A}_i(Z)$ stands for the mean value of the mass number of
%% fragments with $Z$ in the reaction system $i$.  We have observed in
%% the simulation results \cite{ONOk} that the linearity is very well
%% satisfied when the system $i$ is varied, which is an evidence of the
%% validity of the form Eq.\ (\ref{eq:KNZfit}).  The $Z$-dependence of
%% the l.h.s.\ of Eq.\ (\ref{eq:linearrel}) is shown in Fig.\
%% \ref{fig:logyfit-deltazoasq} by filled squares when we take the
%% reaction systems $\calcium[40]+\calcium[40]$ and
%% $\calcium[60]+\calcium[60]$.  Now we can obtain $\zeta(Z)$ in a
%% different way by the fitting of $K(N,Z)$, which is shown by the solid
%% line if Fig.\ \ref{fig:logyfit-deltazoasq}.  The two results agree
%% very well, including the detailed structure of the $Z$-dependence,
%% which is another evidence of the validity of Eq.\ (\ref{eq:KNZfit}).

\begin{figure}
\includegraphics[width=0.48\textwidth]{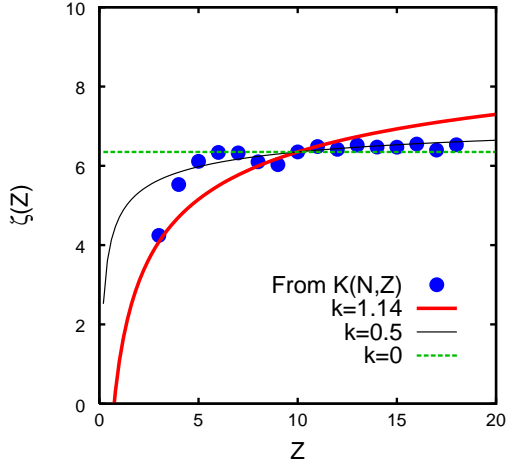}
\caption{\label{fig:logyfit}
  The solid points are the extracted values of the coefficients
  $\zeta(Z)$ of Eq.\ (\ref{eq:KNZfit}) using the combined fragment
  yields of four systems shown in Fig.\ \ref{fig:logyfit-fit}.
  The thick solid curve, the thin sold curve and the dashed line show
  functions $\zeta(Z)\propto 1-k(2Z)^{-1/3}$ normalized at $Z=10$ for
  $k=1.14$, 0.5 and 0, respectively.}
\end{figure}

\begin{figure}
\includegraphics[width=0.48\textwidth]{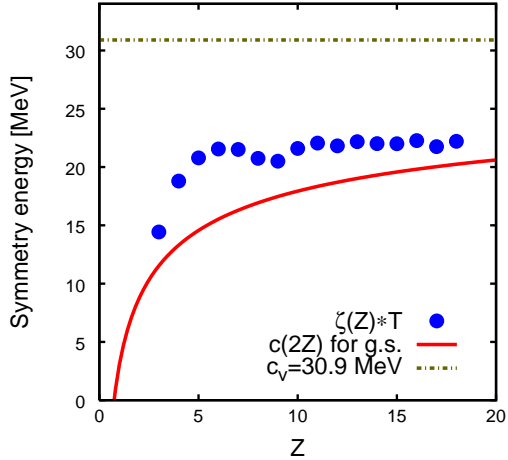}
\caption{\label{fig:logyfit-csym}
  The solid points show $\zeta(Z)T$ when $T=3.4$ MeV is assumed.  The
  dot-dashed horizontal line shows the volume symmetry energy
  $c_\text{v}=30.9$ MeV for the ground state nuclei calculated with
  AMD. The solid line shows the symmetry energy
  $c(A=2Z)=c_\text{v}+c_\text{s}(2Z)^{-1/3}$ for the ground state
  nuclei ($c_\text{v}=30.9$ MeV and $c_\text{s}=-35.2$ MeV).}
\end{figure}

The obtained values of $\zeta(Z)$  are shown in Fig.\
\ref{fig:logyfit} by solid points.  Except for light fragments
($Z\lesssim5$), $\zeta(Z)$ is a smooth function of $Z$ which depends
on $Z$ very weakly.
%% The thin sold line and the
%% thin dashed line in Fig.\ \ref{fig:logyfit} show the results
%% when the reaction systems are limited to only
%% $\calcium[40]+\calcium[40]$ and $\calcium[60]+\calcium[60]$, and to
%% only $\calcium[48]+\calcium[48]$ and $\iron[46]+\iron[46]$,
%% respectively.
Within the uncertainty in the fitting procedure, the trend does not
depend on the number of reaction systems included in Fig.\
\ref{fig:logyfit-fit}.  As expected, the $Z$-dependence becomes more
smooth and stable by including more reaction systems (up to four in
the present work) due to the extension of the range of $N$.

%% The fitting of $B(N,Z)$ by the form of Eq.\
%% (\ref{eq:KNZfit}) is not as good as that of $K(N,Z)$ due to the
%% presence of strong quantum effects such as the paring, shell and
%% $\alpha$-clustering effects in the ground state binding energy.  The
%% extraction of $\zeta(Z)$ for individual $Z$ seems to be meaningful
%% (??) only at the finite temperature where the quantum effects are not
%% strong.

Next we will show that the extracted value of $\zeta(Z)$ can be
regarded as the ratio of the symmetry energy to the temperature.
Under the assumption of equilibrium, $K(N,Z)$ is identified with the
exponent of Eq.\ (\ref{eq:YNZequi}) with inverted signs. If we
approximate the nuclear free energy $G_\text{nuc}(N,Z)$ by a smooth
function of $(N,Z)$, $K(N,Z)$ can be written as
\begin{equation}
K(N,Z)=f(A,Z)+g(A)\frac{(N-Z)^2}{N+Z}+aN+bZ,
\end{equation}
where $f(A,Z)$ and $g(A)$ have smooth $A$ dependence.  Specifically,
we associate the symmetry energy $c(A)$ in $G_\text{nuc}(N,Z)$ with
$g(A)$ by $g(A)=c(A)/T$. $f(A,Z)$ and the linear terms in $N$ and $Z$
can be associated with terms such as volume, surface and Coulomb terms
in the free energy and the chemical potential terms in Eq.\
(\ref{eq:YNZequi}).  Because the range of important $N$ for each $Z$
is limited, we can expand $f(A,Z)$ and $g(A)$ with respect to $N$ for
each $Z$.  By denoting the typical value of $A$ for each $Z$ by
$\bar{A}(Z)$, $f(A,Z)\approx
f(\bar{A}(Z),Z)+f'(\bar{A}(Z),Z)(N+Z-\bar{A}(Z))$ and $g(A)\approx
g(\bar{A}(Z))$.  The higher order terms are negligible as can be
checked directly by assuming the nominal values of the liquid-drop
coefficients.

Thus we have shown that Eq.\ (\ref{eq:KNZfit}) is a natural form
and that the coefficient $\zeta(Z)$ is related to the symmetry
free energy $c(A)$ by
\begin{equation}
\zeta(Z)=c(\bar{A}(Z))/T.
\end{equation}
In principle, $\zeta(Z)$ could depend on $Z$, reflecting the
size-dependence of the symmetry energy $c(A)$.  However, the extracted
values of $\zeta(Z)$ in Fig.\ \ref{fig:logyfit} are almost independent
of $Z$ for $Z\gtrsim5$.  This suggests a reduction of the surface
contribution to the symmetry energy.  The three curves in Fig.\
\ref{fig:logyfit} show functions $\zeta(Z)\propto 1-k(2Z)^{-1/3}$ for
different surface-to-volume ratios $k=1.14$ (thick solid line), 0.5
(thin solid line) and 0 (horizontal dashed line), respectively.  The
curves are normalized at $Z=10$.  The $Z$-dependence of $\zeta(Z)$
cannot be explained by the surface-to-volume ratio
$k=-c_\text{s}/c_\text{v}=1.14$ for the symmetry energy of ground
state nuclei.  The result shows that the surface effect in $\zeta(Z)$
is reduced to between $k=0$ and 0.5.

There can be several possible explanations for the weakening of the
surface symmetry free energy.  First of all, it is not very surprising
that the coefficient in the free energy at a finite temperature is
different from that at the zero temperature.  As is well known, the
surface tension reduces towards zero when the temperature is raised
towards the critical temperature. A similar effect has been obtained
for the symmetry free energy in Thomas-Fermi surface calculations
\cite{LATTIMER,LATTIMERa}.  If we adopt the formula in Ref.\
\cite{LATTIMER}, the reduction factor of the surface symmetry energy
is approximately $[1-(T/T_\text{c})^2]^2=0.91$ for $T=3.4$ MeV (see
below) and the critical temperature $T_\text{c}=16$ MeV.  This
reduction is, however, not sufficient to explain the weak
$Z$-dependence of $\zeta(Z)$.  Another explanation may be associated
with the fact that fragments are not isolated when they are formed.
When the density fluctuation is developing from a uniform low density
matter, the fragments are still interacting with attractive force
through their surfaces.  Therefore, surface free energies could be
expected to be smaller for these fragments than for totally isolated
fragments.  Independent of the physical origin for the weakening of
the surface symmetry free energy, it suggests that the volume
quantity, which is the same as that in the infinite nuclear matter,
can be directly obtained by the analysis of the fragmentation results
even though the produced fragments are not very large.

In the above analysis, only the ratio of the symmetry free energy to
the temperature is obtained.  In the present theoretical approach, it
is possible to get the density and the temperature by studying the
response of the results to a change of the symmetry term in the
effective force.  In Ref.\ \cite{ONOk} we have derived the density
$\rho\sim0.08$ $\textrm{fm}^{-3}$ and the temperature $T\sim3.4$ MeV
for the same reaction systems studied in the present paper.  This
result is consistent to the idea that the fragments are formed in a
low density nuclear matter at a finite temperature.  The solid points
in Fig.\ \ref{fig:logyfit-csym} show $\zeta(Z)T$ so that the extracted
$\zeta(Z)$ can be directly compared to the symmetry energy. The value
of the symmetry energy extracted from fragmentation is significantly
lower than that of the bulk nuclear matter at normal nuclear matter
density, $C_\text{sym}(\rho_0) \approx c_\text{v}=30.9$ MeV shown by the
dot-dashed horizontal line in Fig.\ \ref{fig:logyfit-csym}, and this
must be due to the reduced density, $\rho\sim0.08$
$\textrm{fm}^{-3}<\rho_0$.

The solid line in Fig.\ \ref{fig:logyfit-csym} shows the symmetry
energy $[30.9-35.2(2Z)^{-1/3}]$ MeV that has been obtained by fitting
the ground state binding energies calculated with AMD.  The extracted
symmetry energy is smaller than the ground state symmetry energy
probably because of the greater reduction from the surface
contribution when fragments are formed.

In conclusion, we have analysed AMD simulation results from nuclear
collisions of various nuclei with different neutron-to-proton ratios.
Fragment yields from different reaction systems are combined using the
isoscaling relation. The availability of fragment yields over a wide
range of $N$ and $Z$ allows us to extract the symmetry energy at low
density when fragments are formed.  The results are consistent to the
idea that the fragments are formed in nuclear matter at a low density
and at a finite temperature.  The extracted symmetry energy shows
almost no surface effect in it, which suggests that the property of
infinite nuclear matter can be directly obtained from the information
of fragmentation.

\begin{acknowledgments}
  This work was supported by Japan Society for the Promotion of
  Science and the US National Science Foundation under the U.S.-Japan
  Cooperative Science Program (INT-0124186), by the High Energy
  Accelerator Research Organization (KEK) as a Supercomputer Project,
  and by grants from the US National Science Foundation, PHY-0245009,
  PHY-0070161 and PHY-01-10253, and a Grant-in-Aid for Scientific
  Research from the Japan Ministry of Education, Science and Culture.
  The work was also partially supported by RIKEN as a nuclear theory
  project.
\end{acknowledgments}

\end{document}